\documentclass[preprint,number,sort&compress,12pt]{elsarticle}
\usepackage{epsfig}

\begin{document}

\title{Spectrum of cosmic-ray nucleons, kaon production, and the atmospheric muon
charge ratio}
\author{Thomas K. Gaisser}
\address{Bartol Research Institute and Dept. of Physics and Astronomy\\
University of Delaware, Newark, DE, USA}
\ead{gaisser@bartol.udel.edu}

\begin{abstract}
Interpretation of measurements of the muon charge ratio
in the TeV range depends on the spectra of protons and neutrons
in the primary cosmic radiation
and on the inclusive cross sections for production of 
$\pi^\pm$ and $K^\pm$ in the atmosphere.
Recent measurements of the spectra of cosmic-ray nuclei
are used here to estimate separately the energy spectra of protons and
neutrons and hence to calculate the charge separated hadronic
cascade in the atmosphere.  From the corresponding production spectra
of $\mu^+$ and $\mu^-$ the $\mu^+/\mu^-$ ratio is calculated and
compared to recent measurements.  The comparison leads to
a determination of the relative contribution of kaons and pions.
Implications for the spectra of $\nu_\mu$ and $\bar{\nu}_\mu$
are discussed.
\end{abstract}

\maketitle

\section{Introduction}

The muon charge ratio in the TeV range has been measured
by MINOS~\cite{MINOScharge,Schreiner} and more recently by OPERA~\cite{OPERA}.
Both analyses use an analytic approximation as a framework
for making an inference about the separate contributions of
the pion and kaon channels to the charge asymmetry.  In this
paper a more detailed derivation of the muon charge ratio is
used for the analysis.  The muon charge ratio is expressed in terms
of the spectrum-weighted moments for production of $\pi^\pm$ and
$K^\pm$ by protons and neutrons in the primary cosmic radiation, 
following the analysis of Lipari~\cite{Lipari93}.
The analysis here accounts for the special
contribution of associated production of charged, positive kaons.

This analysis also accounts for the effect of the energy 
dependence of the composition
of the primary cosmic-ray nuclei.  Measurements from 
ATIC~\cite{ATIC} and CREAM~\cite{CREAM,CREAM1} indicate that the spectra
of helium and heavier nuclei become somewhat harder than the 
spectrum of protons above several
hundred GeV.  This feature for helium was recently confirmed
by PAMELA~\cite{PAMELA}.

Because muon neutrinos are produced 
together with muons in the processes
\begin{equation}
\pi^\pm\rightarrow\mu^\pm+\nu_\mu(\bar{\nu}_\mu) \,\,{\rm and}\,\,
K^\pm\rightarrow\mu^\pm+\nu_\mu(\bar{\nu}_\mu),
\end{equation}
 these results also apply to $\nu_\mu$ and $\bar{\nu}_\mu$.
In the TeV range and above the contribution of muon decay
to the intensity of muon neutrinos is negligible.  For
reasons of kinematics, kaons are relatively more important
for neutrinos at high energy than for muons.  An additional goal of
this paper is to draw attention to the implications of the muon
results for atmospheric neutrinos in the TeV energy range and beyond.

\section{Muon charge ratio}

The excess of $\mu^+$ in atmospheric muons can be traced to the
excess of protons over neutrons in the primary cosmic-ray beam coupled
with the steepness of the cosmic-ray spectrum, which emphasizes
the forward fragmentation region in interactions of the incident
cosmic-ray nucleons.  The classic derivation of the muon charge
ratio~\cite{Frazer} considers muon production primarily through the channel
$p\rightarrow \pi^\pm\,+$~anything.  The atmospheric cascade 
equation for the intensity of nucleons as a
function of slant depth $X$ in the atmosphere 
is solved separately for $N = n+p$ and $\Delta = p - n$
subject to the appropriate boundary conditions.  For the total
intensity of nucleons as a function of slant depth $X$~(g/cm$^2$)
\begin{equation}
\phi_N(E)\; =\; \phi_N(0)\times \exp(-{X\over\Lambda_N})
\label{nucleons}
\end{equation}
where the nucleon attenuation length is $\Lambda_N\,=\,\lambda/(1-Z_{NN})$
and $\lambda$ is the interaction length of nucleons in the atmosphere.
The corresponding result for $\Delta(X) = p(X) - n(X)$ is
\begin{equation} 
\Delta(X)\;=\; \delta_0\,\phi_N(0)\times \exp(-{X\over \Lambda_-}),
\label{difference}
\end{equation}
where
\begin{equation}
\delta_0\,=\,{p(0) - n(0)\over p(0) + n(0)}\;\;{\rm and}\;\;\;
{1\over\Lambda_-}\,=\,{1-Z_{pp}+Z_{pn}\over1+Z_{pp}+Z_{pn}}\,{1\over\Lambda_N}.
\label{definition1}
\end{equation}

The $Z$-factors (like $Z_{NN} = Z_{pp}+Z_{pn}$) are spectrum-weighted moments
of the inclusive cross sections for the corresponding hadronic process.
For example, a particularly important moment for this paper is
\begin{equation}
Z_{pK^+}\,=\,{1\over\sigma}\int\,x^\gamma{{\rm d}\sigma(x)\over {\rm d}x} {\rm d}x
\label{ZpKplus}
\end{equation}
for the process
\begin{equation}
 p + air \rightarrow K^+ \,+\,\Lambda\,+\;{\rm anything}.
 \label{associated}
\end{equation}
The normalized inclusive cross section is weighted by $x^\gamma$
where $\gamma$ is the integral spectral index for a power-law
spectrum and $x = E_K/E_p$.  Feynman scaling is assumed in these approximate
formulas, so the parameters may vary slowly with energy, especially
near threshold.  However, the scaling approximation is relatively good because the
moment weights the forward fragmentation region.

\subsection{Charged pion channel}

The next step is to solve the coupled equations for the production
of charged pions by nucleons separately for $\Pi^+(X)+\Pi^-(X)$ and
for $\Delta_\pi\,=\,\Pi^+(X)-\Pi^-(X)$.  The solutions are then convolved
with the probability per g/cm$^2$ for decay to obtain the corresponding
production spectra of muons and neutrinos.  The decay kinematic factors are
\begin{equation} 
{1-r_\pi^{\gamma+1}\over(\gamma+1)(1-r_\pi)}\;\;{\rm and}\;\;\;{\epsilon_\pi\over \cos\theta E_\mu}
\,{1-r_\pi^{\gamma+2}\over(\gamma+2)(1-r_\pi)}
\label{muons}
\end{equation}
for muons and
\begin{equation}
{(1-r_\pi)^\gamma\over(\gamma + 1)}\;\;{\rm and}\;\;\;{\epsilon_\pi\over \cos\theta E_\mu}\,
{(1-r_\pi)^{(\gamma+1)}\over(\gamma+2)}
\label{neutrinos}
\end{equation}
for neutrinos.  In each of Eqs.~\ref{muons} and~\ref{neutrinos} the
first expression is a low-energy limit and the second a high energy limit,
where low and high are with respect to the critical energy $\epsilon_\pi$.
The ratio $r_\pi = m_\mu^2/m_\pi^2 = 0.5731$.  The forms for two-body decay
of charged kaons are the same with $r_K=0.0458$.

The production spectra are then integrated over slant depth through the
atmosphere to obtain the corresponding contributions to the lepton fluxes.  Finally, 
the low and high-energy forms are combined into a single approximate expression.

For example, for the flux of $\nu_\mu+\bar{\nu}_\mu$ the expression is
\begin{eqnarray}
\phi_\nu(E_\nu)& = & \phi_N(E_\nu) \nonumber \\
 & \times & \left\{{A_{\pi\nu}\over 1 + 
B_{\pi\nu}\cos(\theta)E_\nu / \epsilon_\pi}
\,+\,{A_{K\nu}\over 1+B_{K\nu}\cos(\theta)E_\nu / \epsilon_K}\right.\nonumber \\
& & \left. +\,\,\,{A_{{\rm charm}\,\nu}\over 1+B_{{\rm charm}\,\nu}\cos(\theta)E_\nu / \epsilon_{\rm charm}}\right\}.
\label{angular}
\end{eqnarray}
Here $\phi_N(E_\nu) = dN/d\ln(E_\nu)$ is the primary spectrum
of nucleons ($N$) evaluated at the energy of the neutrino.
The three terms in brackets correspond to production from leptonic
and semi-leptonic decays of pions, kaons and charmed hadrons respectively.  
The term for prompt neutrinos from decay of charm has been included in Eq.~\ref{angular}
(see Ref.~\cite{DesiatiGaisser})
but will not be discussed further here.

The numerator of each term of Eq.~\ref{angular} has the form
\begin{equation}
A_{i\nu}\;=\;{Z_{Ni}\times BR_{i\nu} \times Z_{i\nu} \over 1-Z_{NN}}
\label{numerator}
\end{equation}
with $i\,=\,\pi^\pm,\,K,\,{\rm charm}$ and $BR_{i\nu}$ is the branching ratio
for $i\rightarrow\nu$.   The first $Z$-factor in the numerator 
is the spectrum weighted moment
of the cross section for a nucleon (N) to produce a secondary hadron $i$
from a target nucleus in the atmosphere, defined as in Eq.~\ref{ZpKplus}.
The second $Z$-factor is the
corresponding moment of the decay distribution for $i\rightarrow \nu + X$,
which is written explicitly in Eq.~\ref{neutrinos}.  The second term in
each denominator is the ratio of the low-energy to the high-energy form
of the decay distribution~\cite{Gaisser}.
The forms for muons are the same, but the kinematic factors differ
in a significant way (Eq.~\ref{muons} instead of Eq.~\ref{neutrinos}).
Explicitly, for neutrinos
\begin{equation}
B_{\pi\nu}\,=\,\left({\gamma + 2\over \gamma+1}\right)\,\left({1\over 1-r_\pi}\right)\,
\left({\Lambda_\pi - \Lambda_N\over \Lambda_\pi\ln(\Lambda_\pi/\Lambda_N)}\right)
\label{Bnufactor}
\end{equation}
and for muons
\begin{equation}
B_{\pi\mu}\,=\, \left({\gamma + 2\over \gamma+1}\right)\,\left({1-(r_\pi)^{\gamma+1}\over 1-(r_\pi)^{\gamma+2}}\right)\,
\left({\Lambda_\pi - \Lambda_N\over \Lambda_\pi\ln(\Lambda_\pi/\Lambda_N)}\right).
\label{Bfactor}
\end{equation}
The forms for kaons are the same as functions of $r_K$ and $\Lambda_K$.

The separate solutions for $\pi^+\rightarrow \mu^+\,+\,\nu_\mu$
and $\pi^-\rightarrow \mu^-\,+\,\bar{\nu}_\mu$ have the form
\begin{equation}
\phi_\pi(E_\mu)^\pm = \phi_N(E_\mu){A_{\pi\mu}\times0.5(1\pm\alpha_\pi\beta\delta_0)\over
1+B_{\pi\mu}^\pm \,
\cos(\theta)E_\mu/\epsilon_\pi},
\label{muplusminus}
\end{equation}
where
$$B_{\pi\mu}^\pm\,=\,B_{\pi\mu}{1\pm\alpha_\pi\beta\delta_0\over
                          1\pm c_\pi\alpha_\pi\beta\delta_0}.
                          $$
Here $$\beta = {1 - Z_{pp}-Z_{pn}\over 1-Z_{pp}+Z_{pn}}\approx 0.909;\,\,
\beta_\pi\,=\,{1-Z_{\pi^+\pi^+}-Z_{\pi^+\pi^-}\over 1-Z_{\pi^+\pi^+}+Z_{\pi^+\pi^-}}\approx 0.929;$$
$$\alpha_\pi = {Z_{p\pi^+} - Z_{p\pi^-}\over Z_{p\pi^+}+Z_{p\pi^-}}\approx 0.165$$ 
and
$$c_\pi= {1-\Lambda_N/\Lambda_\pi\over1-\beta\Lambda_N/(\beta_\pi\Lambda_\pi)}
\left[1+{ln(\beta_\pi/\beta)\over ln(\Lambda_\pi/\Lambda_N)}\right]\approx 1.01.$$
The numerical values are based on fixed target data in the energy range of hundreds
of GeV~\cite{Gaisser}.
The factors $B_{\pi\mu}^\pm$ differ by less than one per cent.  To this accuracy,
the charge ratio of muons can therefore be written in the form
\begin{equation}
{\mu^+\over \mu^-}\,\approx\,{1\,+\,\beta\delta_0\alpha_\pi\over 1\,-\,\beta\delta_0\alpha_\pi}\,=
\,{f_{\pi^+}\over 1\,-\,f_{\pi^+}},
\label{muratio}
\end{equation}
where $f_{\pi^+}\,=\,(1\,+\,\beta\delta_0\alpha_\pi)/2$ is the
fraction of positive muons from decay of charged pions.  

\subsection{Leptons from decay of kaons}

The situation becomes more complex when the contribution from kaons is considered.
In the first place, because the critical energies are significantly different
for pions and kaons, the two contributions have to be followed separately.
In addition the charge ratio of muons from decay of charged kaons
is larger than that from pion decay because the process of associated production
in Eq.~\ref{associated} has no analog for forward
production of $K^-$.  Instead, associated production by neutrons
leads to $\Lambda\,\bar{K^0}$.

For the charge separated analysis of kaons it is useful to divide
kaon production by nucleons into a part in which
 $K^+$ and $K^-$ are produced equally by neutrons and by protons
 and another for associated production,
which is treated separately.  Then in the approximation that kaon production
by pions in the cascade is neglected, the spectrum of negative muons
from decay of $K^-$ is
\begin{equation}
\phi_K(E_\mu)^- = {Z_{NK^-}\over Z_{NK}}\,\phi_N(E_\mu)\,
{A_{NK}\over 1 + B_{K\mu}\cos(\theta) E_\mu/\epsilon_K}.
\label{Kminus}
\end{equation}

There is an equal contribution of central production to positive kaons, but
in addition there is the contribution from associated production.  The total contribution of the
kaon channel to positive muons is
\begin{equation}
\phi_K(E_\mu)^+ \,=\, \phi_N(E_\mu)\,A_{NK}\times {{1\over2}(1+\alpha_K\beta\delta_0)\over
  1\,+\,B_{K\mu}^+\cos(\theta) E_\mu/\epsilon_K}.
\label{Kplus}
\end{equation}
Here $$\alpha_K\,=\,{Z_{pK^+}\,-\,Z_{pK^-}\over Z_{pK^+}\,+\,Z_{pK^-}}$$
and
$$B_{K\mu}^+\,=\,B_{K\mu}\times {1+\beta\delta_0\alpha_K\over
1+\beta\delta_0\alpha_K(1-ln(\beta)/ln(\Lambda_K/\Lambda_N ) )}.$$

Combining the expressions for $\mu^+$ and $\mu^-$ from 
pions (Eq.~\ref{muplusminus}) and from kaons (Eqs.~\ref{Kminus}
and~\ref{Kplus}), the muon charge ratio is
\begin{eqnarray}
{\mu^+\over \mu^-}&=&\left[{f_{\pi^+}\over 1 \,+\, B_{\pi\mu}\,\cos(\theta) E_\mu/\epsilon_\pi}
+\,{{1\over2}(1+\alpha_K\beta\delta_0)\,A_{K\mu}/A_{\pi\mu}\over 1\, +\, B_{K\mu}^+\,\cos(\theta) E_\mu/\epsilon_K}\right] 
\nonumber \\
& &\times \left[{(1-f_{\pi^+})\over 1 \,+\, B_{\pi\mu}\,\cos(\theta) E_\mu/\epsilon_\pi}
+\,{(Z_{NK^-}/Z_{NK})\,A_{K\mu}/A_{\pi\mu}\over 1\, +\, B_{K\mu}\,\cos(\theta) E_\mu/\epsilon_K}\right]^{-1}.
\label{muchargeratio}
\end{eqnarray}
For the pion contribution, isospin symmetry allows the pion terms in the numerator and
denominator to be expressed in terms of $f_\pi^+$ as defined after Eq.~\ref{muratio} above.
The kaon contribution does not have the same symmetry. 
Numerically, however, the differences are at the level of a few per cent, as 
discussed in the results section.

\section{Primary spectrum of nucleons}

What is relevant for calculating the inclusive spectrum of leptons in the atmosphere is
the spectrum of nucleons per GeV/nucleon.  This is because, to a good approximation,
the production of pions and kaons occurs at the level of collisions 
between individual nucleons in the colliding nuclei.
To obtain the composition from which the spectrum of nucleons can be
derived we use the measurements of CREAM~\cite{CREAM,CREAM1}, grouping their
measurements into the conventional five groups of nuclei, H, He, CNO, Mg-Si and Mn-Fe.
 
\begin{figure}[thb]
\begin{center}
\epsfig{file=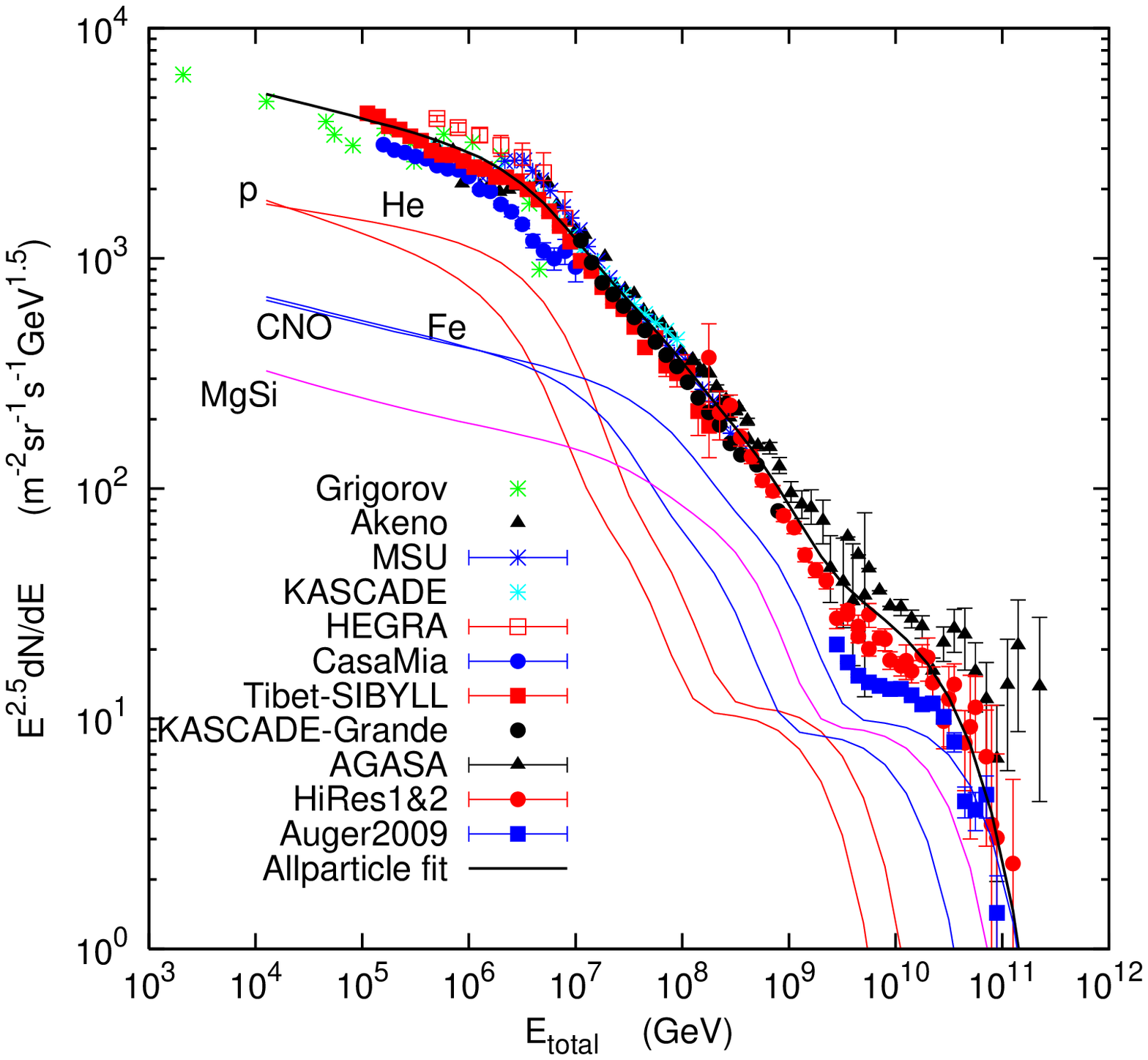,width=3.1in}\epsfig{file=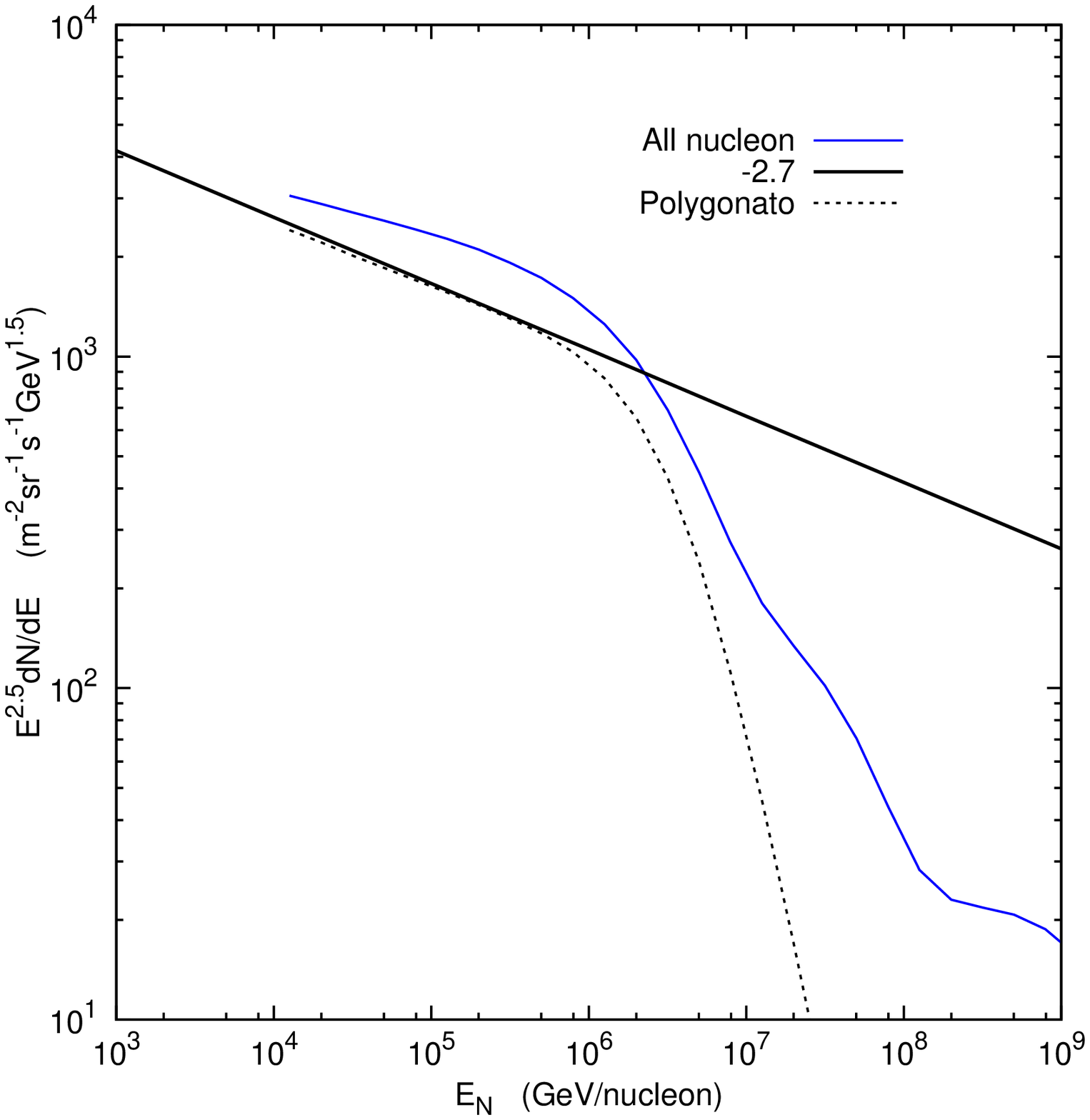,width=2.5in}
\caption{Left: three-population model of the cosmic-ray spectrum from
Eq.~\protect\ref{ModelH3} compared to data~\protect\cite{Grigorov,Akeno,MSU,Antoni,HEGRA,CasaMia,Tibet,KG,AGASA,HiRes,Auger}.
  The extra-galactic 
population in this model has a mixed composition.  Right: 
Corresponding fluxes of nucleons
compared to an $E^{-2.7}$ differential spectrum
of nucleons and to the all nucleon flux implied
by the Polygonato model (galactic component only)~\protect\cite{Polygonato}.  
}
\label{fig1}
\end{center}
\end{figure}

Direct measurements of primary nuclei extend only to $\sim100$~TeV total energy.
Because we want to calculate spectra of muons and neutrinos up to a PeV, we need to extrapolate 
the direct measurements to high energy in a manner that is consistent with measurements of
the all-particle spectrum by air shower experiments in the knee region (several PeV) and beyond,
as illustrated in the left panel of Fig.~\ref{fig1}.
To do this we adopt the proposal of Hillas~\cite{Hillas} to assume three populations of cosmic rays.
The first population can be associated with acceleration by supernova remnants, with the
knee signaling the cutoff of this population.  The second population is a higher-energy
galactic component of unknown origin (``Component B"), while the highest energy population
is assumed to be of extra-galactic origin.

Following Peters~\cite{Peters} 
we assume throughout that the knee and other features of the
primary spectrum depend on magnetic rigidity,
\begin{equation}
R\;=\;{pc\over Ze}, 
\label{rigidity}
\end{equation}
where $Ze$ is the charge of a nucleus of total energy
$E_{\rm tot} = pc$.  The motivation is that both acceleration
and propagation in models that involve collisionless diffusion
in magnetized plasmas depend only on rigidity.  The rigidity determines
the gyroradius of a particle in a given magnetic field $B$
according to
\begin{equation}
r_L\;=\;R\,/\,B\,.
\label{gyroradius}
\end{equation}

\begin{table}[t!]
\begin{center}
\begin{tabular}{r|r|ccccc}  \hline
$R_c$ & $\gamma$&  p & He & CNO & Mg-Si & Fe \\ \hline  \hline
$\gamma$ for Pop. 1 & ---- & 1.66 & 1.58 & 1.63 & 1.67 & 1.63 \\ \hline
Population 1: $ 4$ PV & see line 1 & 7860  &   3550   &   2200     &   1430 & 2120  \\ \hline
Pop. 2: $30$ PV & 1.4 &20  &   20     &   13.4     &   13.4 & 13.4  \\
Pop. 3 (mixed): $2$ EV &1.4 & 1.7  &   1.7     &   1.14     &   1.14 & 1.14  \\ 
    " (proton only): $60$ EV & 1.6 & 200. & 0 & 0 & 0 & 0 
\end{tabular}
\caption{Cutoffs, integral spectral indices and normalizations constants $a_{i,j}$
for Eq.~\ref{ModelH3}.}
\label{tab1}
\end{center}
\end{table}

Peters pointed out that if there is a characteristic rigidity, $R_c$
above which a particular acceleration process reaches a limit
(for example because the gyroradius is larger that the accelerator),
then the feature will show up in total energy first for protons, then for
helium and so forth for heavier nuclei according to
\begin{equation}
E^c_{tot} \;=\;A\times E_{N,c}\;=\; Ze\times R_c.
\label{Etot}
\end{equation}
Here $E_N$ is energy per nucleon, $A$ is atomic mass and $Ze$ the nuclear charge.
The first evidence for such a Peters cycle associated with the knee of
the cosmic-ray spectrum comes from the unfolding analysis of
measurements of the ratio of low-energy muons to electrons at the
sea level with the KASCADE detector~\cite{Antoni}.

In what follows we assume that each of the three components ($j$)
contains all five groups of nuclei and cuts off
exponentially at a characteristic rigidity $R_{c,j}$.  Thus the all-particle
spectrum is given by
\begin{equation}
\phi_i(E)\;=\;\Sigma_{j=1}^3\,a_{i,j}\,
E^{-\gamma_{i,j}}\times \exp\left[-{E\over Z_i R_{c,j}}\right].
\label{ModelH3}
\end{equation}
The spectral indices for each group and the normalizations are given explicitly in
Table~\ref{tab1}.  The parameters for Population 1 are from Refs.~\cite{CREAM,CREAM1},
which we assume can be extrapolated to a rigidity of $4$~PV to describe the knee.
In Eq.~\ref{ModelH3} $\phi_i$ is d$N/$d$\ln E$ and $\gamma_i$ is the integral spectral index.
The subscript $i=1,5$ runs over the standard five groups (p, He, CNO, Mg-Si and Fe), and the 
all-particle spectrum is the sum of the five.

\begin{figure}[thb]
\begin{center}
\epsfig{file=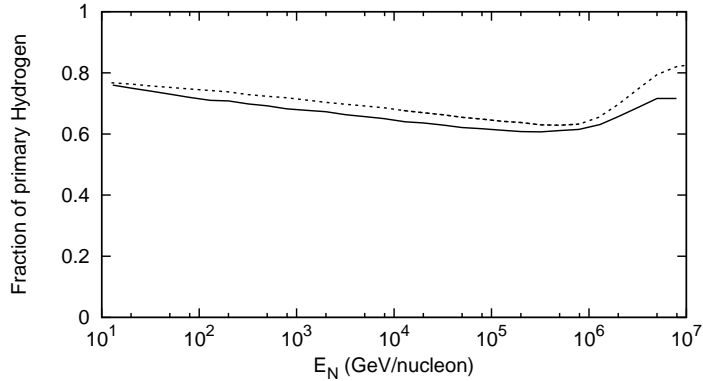,width=3.8in}
\caption{Solid line: charge ratio parameter $\delta_0$ for the model 
with parameters of Table~\protect\ref{tab1}. Dashed line: same for Polygonato
model~\protect\cite{Polygonato}.
}
\label{delta0}
\end{center}
\end{figure}

The composite spectrum corresponding to Eq.~\ref{ModelH3} and Table~\ref{tab1}
is superimposed on a collection of data in the left panel of Fig.~\ref{fig1}.
No effects of propagation in the galaxy or through the microwave background
have been included in this phenomenological model.  For the two galactic
components, however, a consistent interpretation could be obtained with source spectra
$\gamma^*\sim 1.3$ for population 1 and $\gamma^*\sim1.07$ for population 2 together 
with an energy dependent diffusion coefficient $D\sim E^\delta$ with $\delta = 0.33$
for both components to give local spectra of $\gamma = \gamma^*+\delta$ of
$\sim 1.63$ and $~\sim1.4$ respectively.  The extragalactic component comes in
above the energy region of interest for this paper.  We do not discuss it further 
here except to note that the last line of Table~\ref{tab1} gives the parameters
for an extragalactic component of protons only.

The spectrum of nucleons corresponding to Eq.~\ref{ModelH3} is given by
\begin{equation}
\phi_{i,N}(E_N)\;=\;A\times\phi_i(A\,E_N)
\label{EperN}
\end{equation}
for each component and then summing over all five components.
The nucleon spectrum is shown in the right panel of Fig.~\ref{fig1}.

The energy-dependent charge ratio $\delta_0(E_N)$ needed to calculate
the muon charge ratio follows from Eq.~\ref{EperN} and Table~\ref{tab1}.
To a good approximation, it is given by the fraction of free 
hydrogen in the spectrum of nucleons, as shown in Fig.~\ref{delta0}.
The fraction decreases slowly from its low energy value of 
$0.76$ at 10 GeV/nucleon~\cite{PDG} to a minimum of $0.63$ at 300 TeV
and then increases somewhat at the knee.  Note that, because of the
relation among $E_{\rm tot}$, $E_N$ and $R_c$ in Eq.~\ref{Etot}, the steepening
at the knee occurs for nuclei at $Z/A\approx {1\over 2}$ the energy per nucleon
as compared to protons.  Hence the free proton fraction rises again at the knee.

Also shown for comparison in Fig.~\ref{delta0} by the broken line is the $\delta_0$ parameter
for the rigidity-dependent version of the Polygonato model, which has a common
change of slope $\Delta\gamma = 1.9$ at the knee~\cite{Polygonato}.  This gives rise
to the sharp cutoff in the spectrum of nucleons for this model in the right panel of Fig.~\ref{fig1}.  
This version of the Polygonato model is meant to describe only the knee of the
spectrum and the galactic component of the cosmic radiation.  The behavior of the
primary spectrum for $E_N>10^5$~GeV/nucleon does not affect the charge
ratio, which is measured only for $E_\mu < 10^4$~GeV.  It is therefore possible
to consider the difference between the two versions of $\delta_0$ in Fig.~\ref{delta0}
as a systematic effect of the primary composition.

\section{Comparison with data}

We now wish to compare the calculation of
Eq.~\ref{muchargeratio} to various sets of data using
the energy-dependent primary spectrum of nucleons
(Eq.~\ref{EperN}) with parameters from Table~\ref{tab1}.
There are two problems in doing so.  First,
expressions for the intensity of protons
and neutrons from Eqs.~\ref{nucleons} and~\ref{difference} 
and the subsequent equations are valid under the assumption
of a power-law spectrum with an energy independent value of $\delta_0$.  
The assumption of a power law with integral spectral index of $-1.7$
is a reasonable approximation over the range of energies below the knee 
because it affects both charges in the same way.
The proton-neutron difference, however, introduces an explicit energy-dependence
into Eq.~\ref{muchargeratio} that must be accounted for.
  We want to consider the energy range
from 10 GeV to PeV over which the composition changes slowly with energy,
as shown in Fig.~\ref{delta0}.  For estimates here
we use the approximation $\delta_0(E_N)\,=\,\delta_0(10\times E_\mu)$.  

\begin{figure}[thb]
\begin{center}
\epsfig{file=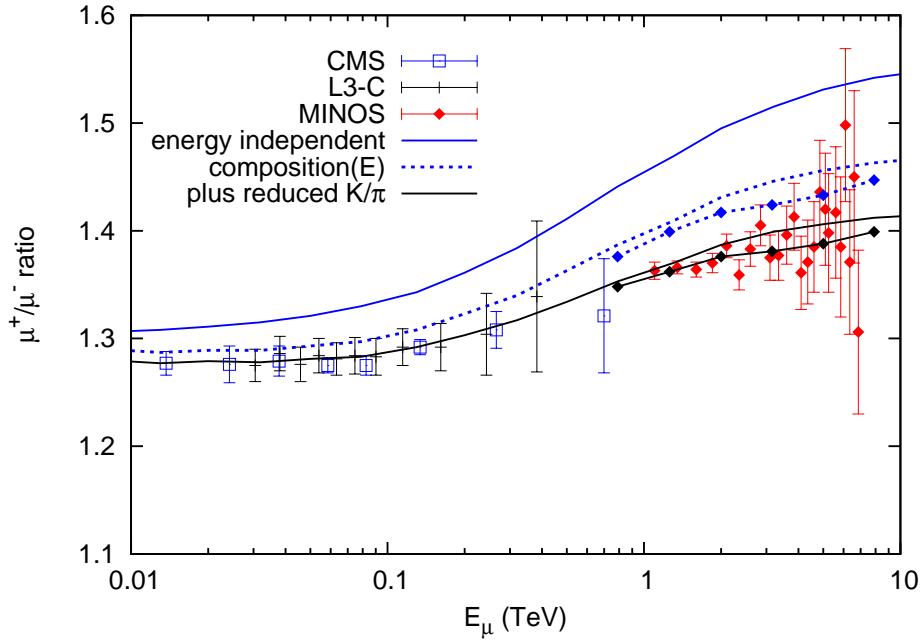,width=5in}
\caption{Muon charge ratio compared to data of CMS~\protect\cite{CMS} and
L3-C~\protect\cite{L3C} below 1 TeV and to MINOS~\protect\cite{MINOScharge}
at higher energy.  The L3-C data plotted here are averaged over
 $0.9\le\cos(\theta)\le 1.0$ for comparison with the calculation for vertical muons.
See text for a description of the lines.}
\label{fig3}
\end{center}
\end{figure}

The other problem is that the data are obtained over a large range of zenith angles, 
and the charge ratio also depends on angle.  The first MINOS publication~\cite{MINOScharge} 
gives $\mu^+/\mu^-$ as a function of the energy of the muon at the surface.  
These data are shown in Fig.~\ref{fig3} along with older high energy data
from the Park City Mine in Utah~\cite{Utah} and data at lower energy from
L3~\cite{L3C} and CMS~\cite{CMS}.  The figure
shows three calculations of the muon charge ratio in the vertical direction
that follow from Eq.~\ref{muchargeratio}.  The highest curve assumes a constant 
composition fixed at its low energy value, $\delta_0\,=\,0.76$~\cite{PDG}.  The middle
curve is the result assuming the energy-dependent composition parameter $\delta_0(E_N)$ 
that corresponds to the parameterization of Table~\ref{tab1} (solid line in Fig.~\ref{delta0}), 
which is still higher than the data.  Both the higher lines assume the nominal
values of the spectrum weighted moments from Ref.~\cite{Gaisser}.  The lowest
curve is obtained by reducing the level of associated production,  by
 changing $Z_{pK^+}$ from its nominal value of $0.0090$ to $0.0079$
 
In order to look for the best fit it is necessary first to account for the
dependence on zenith angle.
The MINOS paper~\cite{MINOScharge} does not give the mean zenith angle
for each energy bin.  However, because of the flat overburden at the Soudan mine
where the MINOS far detector is located, there is a strong correlation between
zenith angle and energy at the surface, as illustrated in Fig.~14 of Ref.~\cite{MINOScharge}.  
Using this
relation we estimate the effective zenith angle as a function of energy from
$\cos(\theta) \approx 0.9$ at 1~TeV to $\cos(\theta)\approx 0.5$ at 7~TeV.
The points connected by lines in the multi-TeV range show the results of the calculation
taking account of the dependence on zenith angle in Eq.~\ref{muchargeratio}.
The lowest set of points has been adjusted to fit the MINOS data by varying the parameter
$Z_{pK^+}$, which reflects $p\rightarrow K^+$.

\begin{figure}[thb]
\begin{center}
\epsfig{file=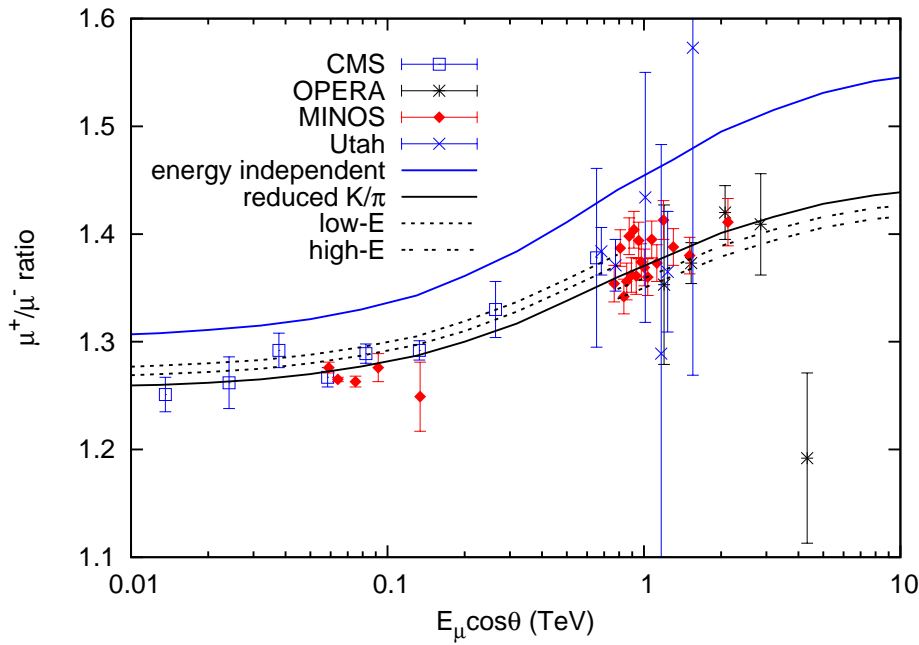,width=5.in}
\caption{Muon charge ratio compared to data of 
CMS~\protect\cite{CMS}, OPERA~\protect\cite{OPERA} and MINOS~\protect\cite{Schreiner}.  
Measurements from the near detector of MINOS~\cite{Jeff} and from Park City~\cite{Utah}.
}
\label{fig4}
\end{center}
\end{figure}

The dependence of $\mu^+/\mu^-$ on zenith angle enters Eq.~\ref{muchargeratio}
in the form $E_\mu\cos(\theta)$.
For this reason the muon charge ratio is often presented as a function of
this combination.  The data of OPERA are presented only in terms of the product
$E_\mu\cos(\theta)$.  Because of the complex overburden at Gran Sasso, there is
no simple relation between zenith angle and energy.  
The MINOS data are also presented in this form in Ref.~\cite{Schreiner}, but the mean
energy for each value of $E_\mu\cos(\theta)$ is not given.  For a primary
cosmic-ray composition that has no energy dependence, Eq.~\ref{muchargeratio}
depends only on $E_\mu\cos(\theta)$.  The effect of the energy-dependence
of composition can threrfore be assessed by comparing the calculation for various
fixed values of $\delta_0$ to the data, which is done in Fig.~\ref{fig4}. 

The upper curve in Fig.~\ref{fig4} is the same as the corresponding curve
in Fig.~\ref{fig3}, plotted for a constant composition with $\delta_0=0.76$,
its value at 10 GeV.
The parameter $\delta_0$ decreases from 0.71 at 100 GeV/nucleon to 0.68 at a TeV,
and from 0.64 at 10 TeV/nucleon to less than 0.62 at 100 TeV.  The full curve through
the data in Fig.~\ref{fig4} is evaluated for $\delta_0=0.665$.  The two broken
lines in the low energy region are plotted for 0.71 and 0.69, while those
at high energy are for 0.64 and 0.62.  A more precise comparison between
data and calculation could be made given complete information about the
distribution of energy within each bin of zenith angle, but it is clear that
the data from the various experiments are reasonably consistent with each
other and with the present calculation.

\section{Summary}

 The muon charge ratio is sensitive both to the proton excess
 in the spectrum of primary cosmic-ray nucleons and to 
 the value of $Z_{pK^+}$.  Using recent data on primary composition,
 we find a proton excess that decreases steadily from 10 GeV/nucleon
 to 500 TeV.  This portion of the cosmic-ray spectrum produces
 muons from a few GeV to well over 10 TeV.
Assuming associated production (Eq.~\ref{associated}) to be the major
uncertainty, a level of associated production in the range $Z_{pK^+} = 0.0079\pm.0002$
is required to fit the observed charge ratio.
For comparison, the nominal value~\cite{Gaisser} is $Z_{pK^+} = 0.0090$.
Keeping the nominal values of all other 
parameters, the fit here corresponds to a ratio
\begin{equation}
R_{K/\pi}\,=\,{Z_{pK^+} + Z_{pK^-}\over Z_{p\pi^+}+Z_{p\pi^-}}\,=\,{0.0079 + 0.0028\over 0.046 + 0.033}
\,=\,0.135.
\label{K2pi}
\end{equation}

It is interesting that analyses of seasonal variations of TeV muons by MINOS~\cite{MINOSseasonal}
and IceCube~\cite{I3seasonal} also suggest a somewhat lower value of $R_{K\pi}$ than its nominal
value of 0.149.  On the other hand, the value in Eq.~\ref{K2pi} still represents a significant
contribution from the $K^+$ decay channel.  If the energy-dependent composition of 
the Polygonato model is used instead, a good fit is obtained with $Z_{pK^+} = 0.0074$,
which reflects the somewhat higher value of $\delta_0$ in the relevant energy range
($0.68$ as compared to $0.64$).  The fraction of kaons would be correspondingly lower
($R_{K/\pi}=0.129$).

\vfill\eject

In the analysis of MINOS, and also in that of OPERA, the muon charge ratio
is written in the form
\begin{eqnarray}
{\mu^+\over \mu^-}&=&\left[{f_{\pi^+}\over 1 \,+\, B_{\pi\mu}\,\cos(\theta) E_\mu/\epsilon_\pi}
+\,{f_{K^+}\,A_{K\mu}/A_{\pi\mu}\over 1\, +\, B_{K\mu}\,\cos(\theta) E_\mu/\epsilon_K}\right] 
\nonumber \\
& &\times \left[{(1-f_{\pi^+})\over 1 \,+\, B_{\pi\mu}\,\cos(\theta) E_\mu/\epsilon_\pi}
+\,{(1-f_{K^+})\,A_{K\mu}/A_{\pi\mu}\over 1\, +\, B_{K\mu}\,\cos(\theta) E_\mu/\epsilon_K}\right]^{-1}
\label{minosratio}
\end{eqnarray}
with $A_{K\mu}/A_{\pi\mu} = 0.054$.  The more correct Eq.~\ref{muchargeratio} has
a different form for the contribution from kaons.
In the MINOS analysis the fitted values of the two free parameters are
$f_\pi^+ = 0.55$ and $f_K^+ = 0.67$.  For $E_\mu\sim$~TeV, $\delta_0\approx 0.64$
for primary energy per nucleon of $10$~TeV.  Thus
$f_\pi^+ = {1\over 2}(1+\alpha_\pi\beta\delta_0) = 0.55$ in agreement with the MINOS analysis.
A precise comparison with the MINOS value for $f_K^+$ is not possible for the
reasons explained after Eq.~\ref{muchargeratio}.  However, numerical differences
are at the level of a few per cent.  For example, $B_{K\mu}^+\approx 0.95\times B_{K\mu}$.
From Eqs.~\ref{Kminus} and ~\ref{Kplus}, the value of 
$f_{K^+} = \phi_K(E_\mu)^+/(\phi_K(E_\mu)^+ + \phi_K(E_\mu)^-) \approx 0.69$
in the TeV region for $\delta_0\approx0.64$.
However, if the expression for the kaon contribution in Eq.~\ref{muchargeratio} is
expressed in terms of $f_K^+$ there is an additional multiplicative factor less than unity.
Thus, although the forms are different, the fits are much the same.

\begin{figure}[thb]
\begin{center}
\epsfig{file=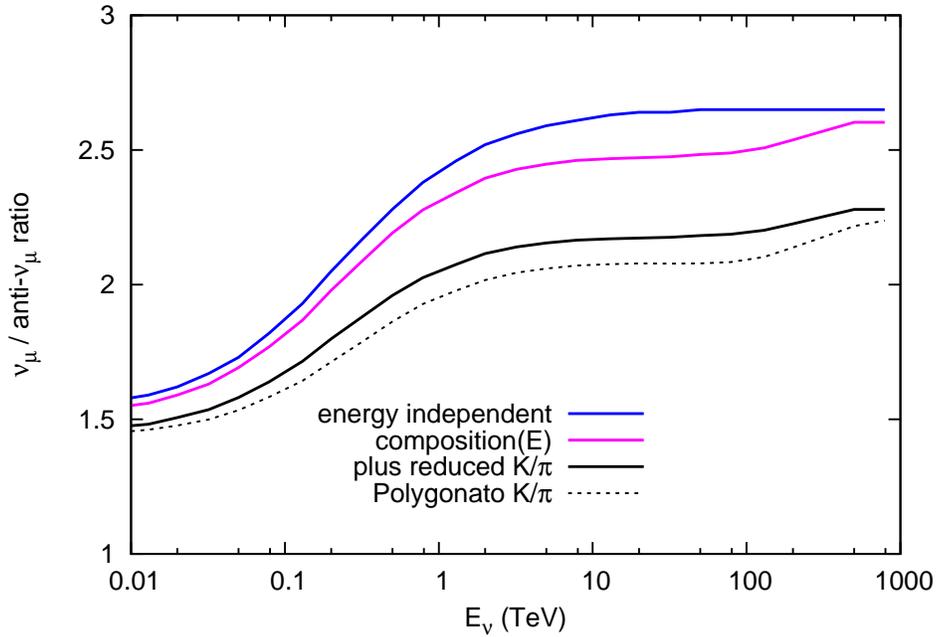,width=5.in}
\caption{ratio of $\nu_\mu/\bar{\nu}_\mu$ calculated with the same parameters
as for the charge ration of muons.  The dashed line shows the results with parameters of
the Polygonato model~\protect\cite{Polygonato}.}
\label{fig5}
\end{center}
\end{figure}

The role of kaons is relatively more important for neutrinos than for muons.  Because
the muon mass is close to that of the pion, the muon carries most of the energy of
the decaying pion.  Kaons split the energy almost equally on average between the 
$\mu$ and the $\nu_\mu$.  The steep spectrum enhances the effect so that kaons
are the dominant source of muon neutrinos above a few hundred GeV.  Forward production
of $K^+$ is therefore particularly important.  The effect is illustrated in 
 Fig.~\ref{fig5},
in which the ratio $\nu_\mu/\bar{\nu}_\mu$ is plotted for the same sequence of assumptions
as in the plot of the muon charge ratio (Fig.~\ref{fig3}).
The implications of the muon charge ratio for neutrinos will be the subject
of a separate paper.

\noindent{\bf Acknowledgments}: I am grateful to Anne Schukraft and Teresa Montaruli
for comments on an early version of this paper.  I am grateful for helpful comments
from an external reviewer and to Jeffrey de Jong and Simone Biagi for information
about MINOS and OPERA.  This research is supported in part
by the U.S. Department of Energy under DE-FG02-91ER40626.

\end{document}